\documentclass[
    ,final            
    ,numberedheadings 
    ,cmfonts          
  ]
  {aipproc}
\layoutstyle{6x9}

\SetInternalRegister\hbadness{8000} 

\begin{document}

\title{Restoration of  U$_A$(1) symmetry and  meson spectrum in hot or dense matter}

\classification{11.10.Wx, 11.30.Rd, 14.40.Aq, 24.85.+p}
\keywords      {NJL model, U$_A$(1) symmetry, Finite temperature and density}

\author{P. Costa}{address={Centro de F\'{\i}sica Te\'{o}rica,
Departamento de F\'{\i}sica, Universidade, P3004-516 Coimbra, Portugal}}

\author{M. C. Ruivo}{address={Centro de F\'{\i}sica Te\'{o}rica,
Departamento de F\'{\i}sica, Universidade, P3004-516 Coimbra, Portugal}}

\author{C. A. de Sousa}{address={Centro de F\'{\i}sica Te\'{o}rica,
Departamento de F\'{\i}sica, Universidade, P3004-516 Coimbra, Portugal}}

\author{Yu.L.~Kalinovsky}{
  address={Universit\'e de  Li\`ege, D\'epartement de  Physique B5, Sart Tilman,B-4000 LIEGE 1, Belgium},
altaddress={Laboratory of Information Technologies, Joint Institute for Nuclear Research, Dubna, Russia \\
E-mail: pcosta@fteor5.fis.uc.pt, maria@teor.fis.uc.pt, celia@teor.fis.uc.pt, kalinov@nusun.jinr.ru}
}

\begin{abstract}
We explore the effects of breaking and restoration of chiral and axial symmetries using an extended  three-flavor Nambu-Jona-Lasinio model that incorporates explicitly the axial anomaly through the 't Hooft interaction. 
We implement a temperature (density) dependence of the anomaly coefficient motivated by lattice results for the topological susceptibility.
The spectrum of scalar and pseudoscalar mesons is analyzed bearing in mind the identification of chiral partners and the study of its convergence.
We also concentrate on the behavior of the mixing angles that give us relevant information on the issue under discussion.
The results suggest that the axial part of the symmetry is restored before the possible restoration of the full U(3)$\otimes$U(3) chiral symmetry might occur.  
\end{abstract}
\maketitle


\section{Introduction}

The study of properties of low-lying hadron spectrum from the viewpoint of QCD dynamics and symmetries has been an intrinsically interesting and relevant topic in physics of strong interactions. In this context, the explicit and spontaneous breaking of chiral symmetry, as well as  the U$_A$(1) anomaly, play a key  role for the generation of the low-lying pseudoscalar meson masses.
The importance for this process of the   U$_A$(1) symmetry breaking, and the consequent violation of the Okubo-Zweig-Iizuka  (OZI) rule, has  been stressed in many phenomenological investigations.  

It is generally expected that ultra-relativistic heavy-ion experiments will provide the strong interaction conditions which will lead to new physics. Restoration of symmetries and deconfinement  are expected to occur under those conditions, allowing for the search of signatures of quark gluon plasma.
In particular, the order of the chiral phase transition and its consequences  for aspects of the dynamical evolution of the system are  important questions.    
It is also believed that at high temperatures the instanton effects are suppressed  due to the Debye-type screening \cite{Gross}. Then an effective restoration of U$_A$(1) symmetry is expected to occur at high temperatures.
In this context, it has been argued that the mass of the $\eta^\prime$ excitation in hot and dense matter should be small, being expected the return of this "prodigal Goldstone boson" \cite{Kapusta}.

There should be other indications of the restoration of the axial symmetry,
like the vanishing  of the topological susceptibility \cite {lattice}, $\chi$, which, in pure color SU(3) theory, can be linked to the $\eta'$ mass through the Witten-Veneziano formula \cite{Veneziano}. In addition, since the presence of the axial anomaly causes flavor mixing, with the consequent violation of the OZI rule, both for scalar and pseudoscalar mesons, restoration  of axial symmetry should have relevant consequences for the phenomenology of meson mixing angles, leading to the recovering of ideal mixing.

We perform our calculations in the framework of an extended  SU(3) Nambu--Jona-Lasinio model Lagrangian density that includes the 't Hooft determinant:
\begin{eqnarray}
{\mathcal L\,}&=& \bar q\,(\,i\, {\gamma}.\partial-\,\hat m)\,q
+ \frac{g_S}{2}\, \sum_{a=0}^8[\,{(\,\bar q\,\lambda^a\, q\,)}
^2+{(\,\bar q \,i\,\gamma_5\,\lambda^a\, q\,)}^2\,]  \nonumber \\
&+& g_D\,\{\mbox{det}\,[\bar q\,(1+\gamma_5)\,q] +\mbox{det}
\,[\bar q\,(1-\gamma_5)\,q]\}. 
\end{eqnarray}
By using a standard hadronization procedure, an effective  action is obtained, leading to gap equations for the constituent quark masses and to meson propagators from which several observables are calculated \cite{costa}.

In what follows, we will concentrate first on the restoration of the symmetries at zero density and finite temperature, and the manifestations of the restoration on the scalar and pseudoscalar meson observables will be analyzed. 
In chiral models, when the coefficient of the anomaly term in the Lagrangian ($g_D$ in the present case) is constant, it happens that, in spite of the decreasing of the  U$_A$(1) violating quantities, the axial  symmetry is not restored due to the fact that the strange quark condensate does not decrease enough. However, an effective restoration may be achieved by assuming that the strength of the anomaly coefficient is  a dropping function of the temperature \cite{alkofer,Bielich,Ohta}.
The analysis  of the temperature dependence of the mixing angles, allowing for the understanding of the evolution of meson quark content and, in particular,  of the role of the strange order parameter in SU(3) chiral partners like $(\sigma,\eta)$ and $(f_0,\eta^\prime)$, can provide further indication of the restoration of the axial symmetry.

Preliminary results from  calculations on lattice QCD at finite chemical potential \cite{Alles} motivates also the study of the restoration of the U$_A$(1) symmetry at finite density. 
  
We study the effective restoration of chiral and axial symmetries with temperature and zero density in Sec. \ref {FTC} and, with finite density and zero temperature in Sec.  \ref{FDC}. Finally, we summarize our conclusions in Sect. \ref {Conc}.     


\section{Finite temperature and zero density case}
\label{FTC}

Restoration of the axial symmetry should be manifest when the effects of anomaly are no longer visible and, in  the present work, we model this feature  by means of a specific temperature dependence of $g_D$.
So, following the methodology of \cite{Ohta}, we  extract the temperature dependence of the anomaly coefficient $g_D$  from the lattice results for the topological susceptibility  \cite{lattice} (see Fig. 1, left panel).

\begin{figure}[t]
\hspace{0.3cm}\includegraphics[width=7cm,height=6.7cm]{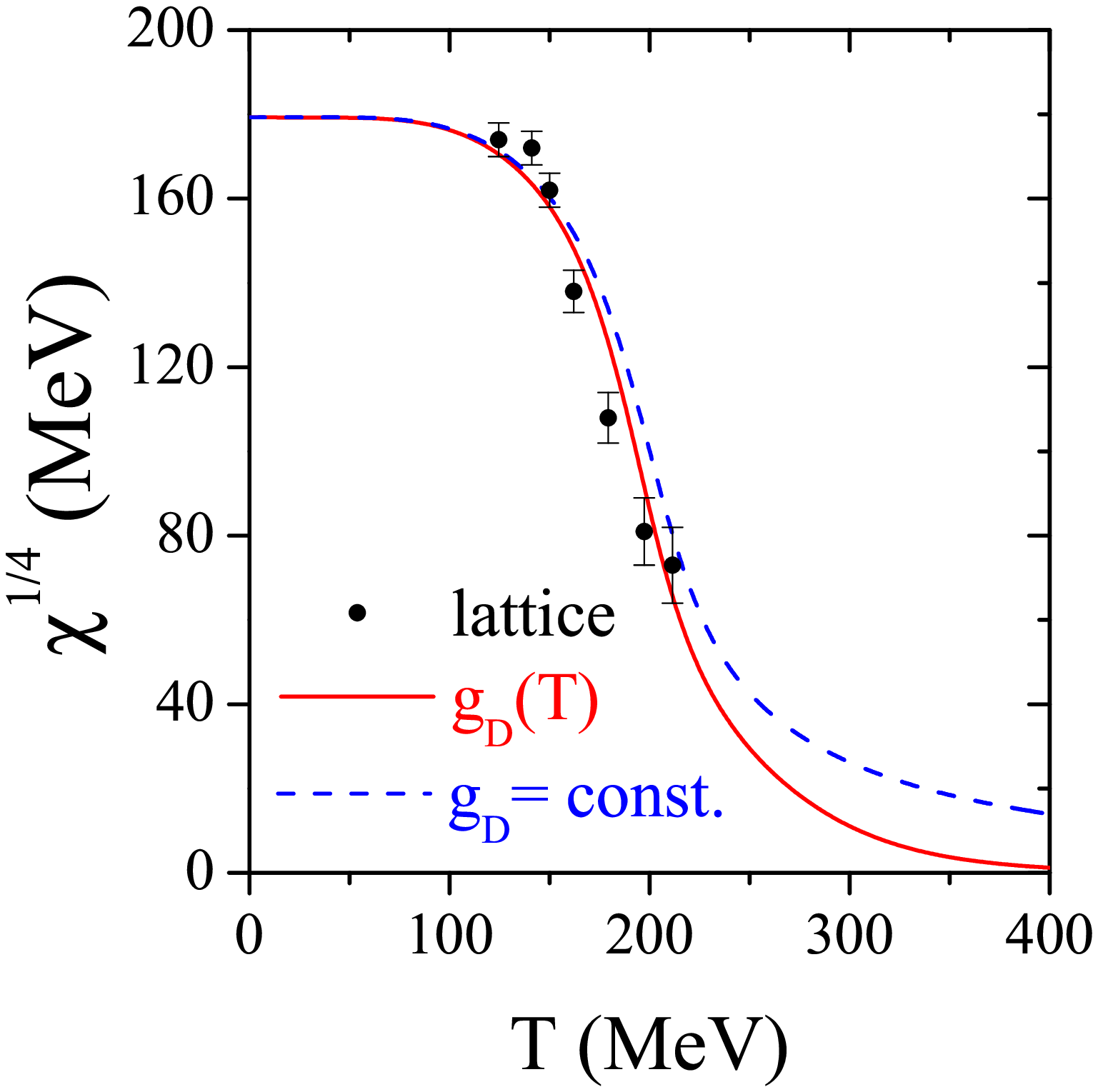}
\hspace{-1cm}\includegraphics[width=10cm,height=7cm]{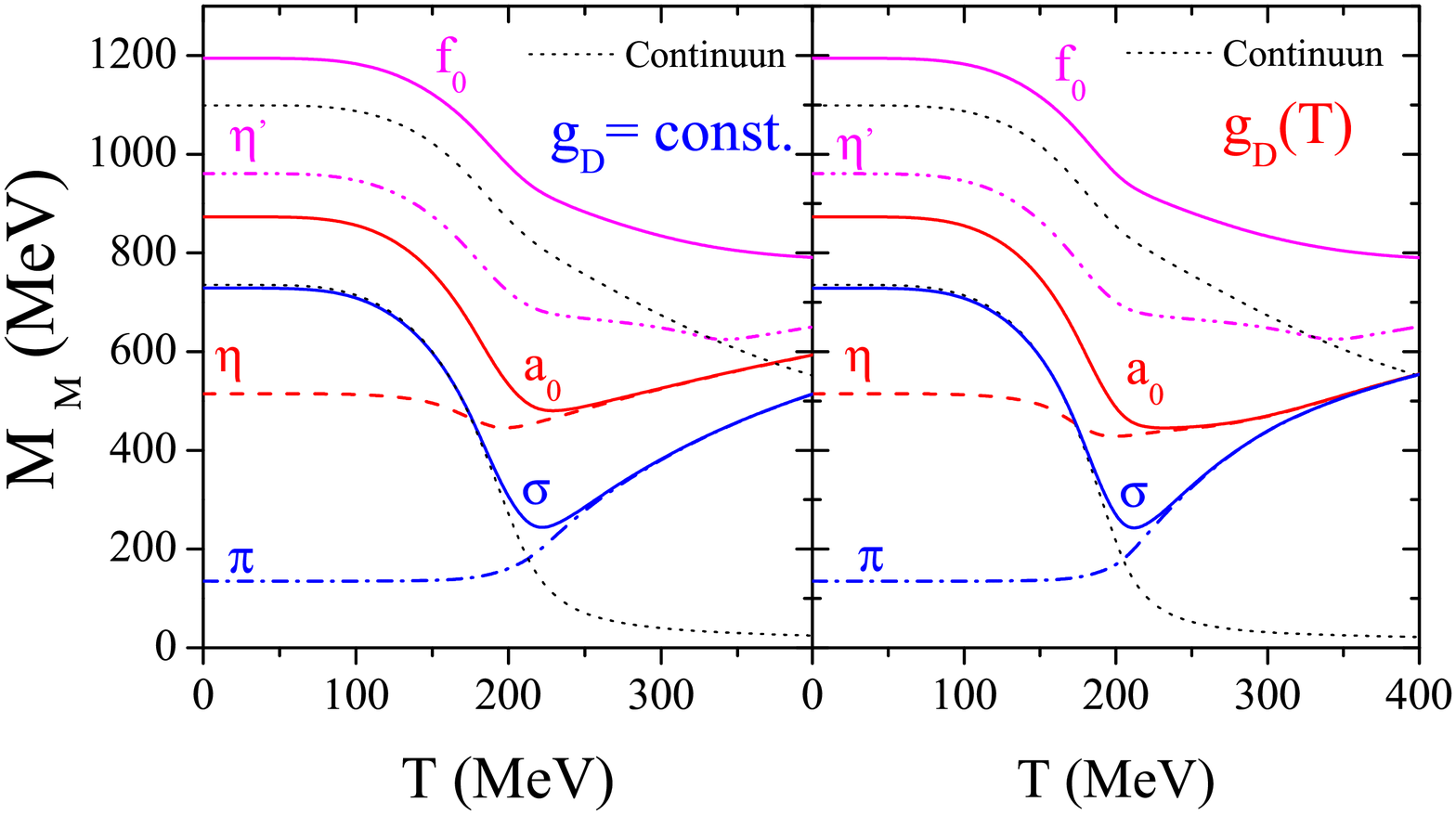}
\caption{Topological susceptibility (left panel): from lattice data plotted with error bars \cite{lattice}; the solid (dashed) line represents our fitting with constant (temperature dependent) $g_D$. Meson masses, as functions of the temperature, with $g_D$ constant (middle panel) and $g_D (T)$ (right panel). The dotted lines indicate the continuum thresholds $2M_u$ and $2M_s$.}
\label{fig:temp}
\end{figure}

Our results, concerning the restoration of the chiral phase transition, at zero density and finite temperature, indicate, as usual, a smooth crossover: at temperatures around $T\approx200$ MeV the mass of the light quarks drops to the current quark mass. The strange quark mass also starts to decrease significantly in this temperature range, however even at $T = 400$ MeV it is still 2 times the strange current quark mass. 
In fact, as $m_u=m_d<m_s$, the (sub)group SU(2)$\otimes$SU(2) is a much better symmetry of the Lagrangian (1). 
So, the effective restoration of the above symmetry  implies the degeneracy between the  chiral partners $(\pi,\sigma)$ and $(a_0,\eta)$ around $T\approx250$ MeV (see Fig. 1, right panel).
For temperatures  about $T\approx350$ MeV, both $a_0$ and $\sigma$ mesons become degenerate with the $\pi$ and $\eta$ mesons, showing an effective restoration of both chiral and axial symmetries.
In fact, the U$_A$(1) symmetry is effectively restored when the U$_A$(1) violating quantities show a tendency to vanish, which means that the four meson masses  degenerate and the topological susceptibility goes to zero. Without the restoration of U$_A$(1) symmetry, the $a_0$ mass was moved upwards and never met the $\pi$ mass as can be seen in Fig. 1, middle panel. The same argument is valid for $\sigma$ meson comparatively with the $\eta$ meson mass. We remember that the determinantal term acts in an opposite way for the scalar and pseudoscalar mesons.
So, only after the effective restoration of U$_A$(1) symmetry we can recover the SU(3) chiral partners $(\pi,a_0)$ and $(\eta,\sigma)$ which are now all degenerate.
However, the $\eta^\prime$ and $f_0$ masses do not yet show a clear tendency to converge in the region of temperatures studied.

In order to understand this behavior, we also analyze the temperature dependence of the mixing angle (see Fig. 2): $\theta_S$ starts at $16^{\circ}$ and goes, smoothly, to the ideal mixing angle $35.264^{\circ}$ and $\theta_P$ starts at $-5.8^{\circ}$ and goes to the ideal mixing angle $-54.7^{\circ}$. 
This means that flavor mixing no more exists.
In fact, analyzing the behavior of the SU(2) chiral partner ($\eta,a_0$) with the temperature (Fig. 1, right panel), we found that the $a_0$ meson is always a purely non-strange $q \bar q$  system while the $\eta$ meson, at $T = 0$ MeV, has a strange component and becomes purely non-strange when $\theta_P$ goes to $-54.7^{\circ}$ at $T \approx 250 $ MeV. At this temperature they start to be degenerate. Concerning the  SU(2) chiral partner ($\pi,\sigma$), at $T = 0$ MeV, $\pi$ is always a light quark system and the $\sigma$ meson has a strange component,  but becomes purely non-strange when $\theta_S$ goes to $35.264^{\circ}$ at $T \approx 250 $ MeV. In summary,  we see that the U$_A$(1) symmetry is effectively restored at $T\approx350$ MeV:   $\pi$, $\sigma$, $\eta$ and $a_0$ mesons become degenerate, the OZI rule is verified and $\chi$ goes asymptotically to zero \cite{costaUA1}.
We remember that  $\pi$, $\sigma$, $\eta_{\rm ns}$ and $a_0$ form a complete representation of U(2)$\otimes$U(2) symmetry, but the $\pi$ and the $a_0$ do not belong to the same SU(2)$\otimes$SU(2) multiplet. 
The partners $(f_0\,,\eta')$, which become purely strange at high temperatures,  do not converge probably due to the fact that chiral symmetry is not restored in the strange sector.

\begin{figure}[t]
\includegraphics[width=0.6\textwidth]{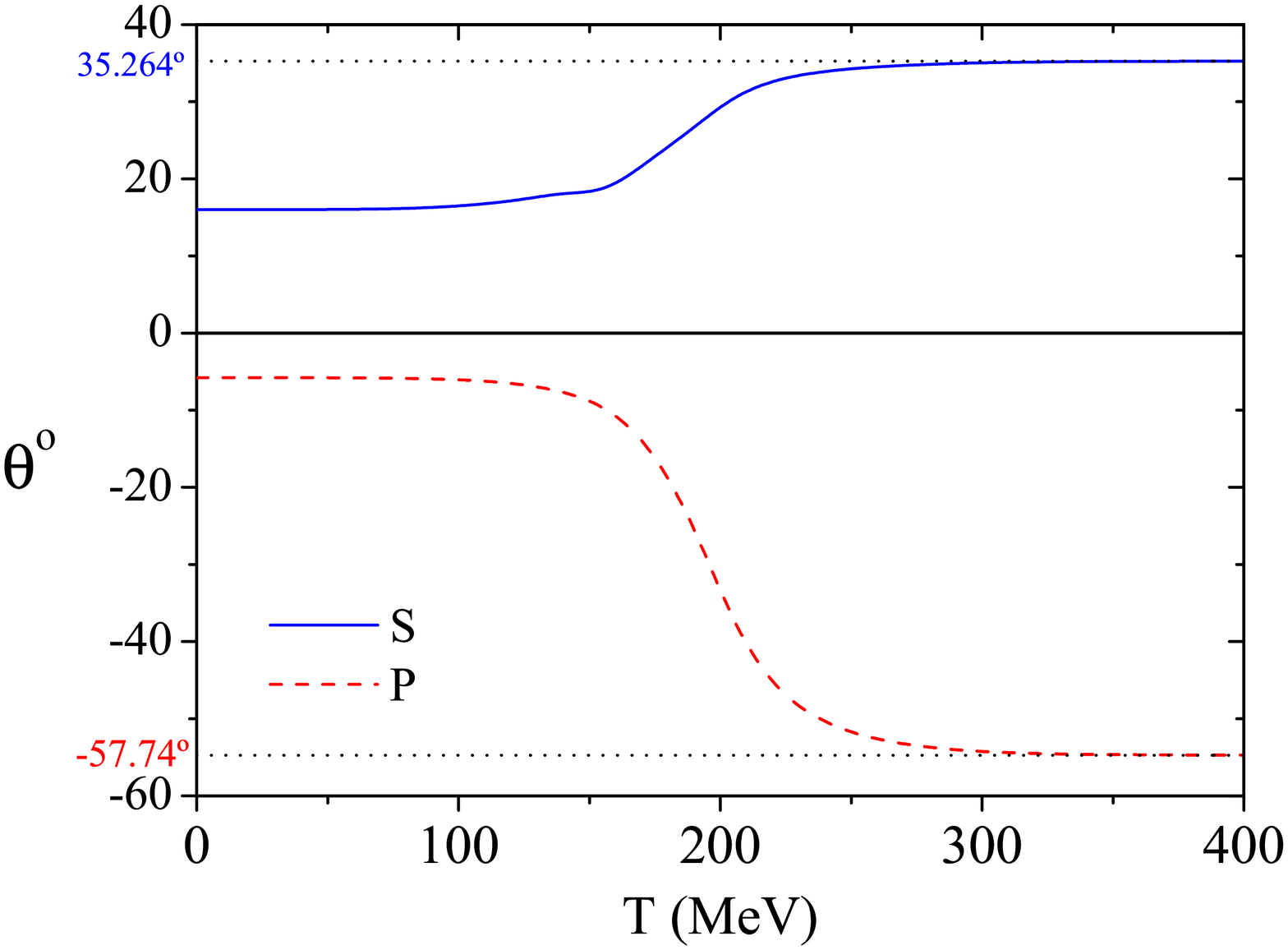}
\caption{The pseudoscalar and scalar mixing angles as functions of the temperature.}
\label{fig:angtemp}
\end{figure}


\section{Finite density and zero temperature case}
\label{FDC}

Recent calculations on lattice QCD at finite chemical potential  motivates also the study of the restoration of the U$_A$(1) symmetry at finite density. There are no firmly lattice results for the density dependence of $\chi$, to be used as input, but the preliminary conclusions of \cite{Alles} suggest a possible decrease of $\chi$. So, it seems reasonable to model the density dependence of $g_D$ extrapolating   from our previous results for the finite temperature case  and proceeding by analogy. Here we present an example (see Fig. 3, left panel) where we consider quark matter simulating "neutron" matter. This "neutron" matter is in $\beta$--equilibrium with charge neutrality, and undergoes a first order phase transition \cite{costa}.
To begin with, we calculate the mixing angles for scalar and pseudoscalar mesons, $\theta_S$ and $\theta_P$, respectively, that are plotted in Fig. 4. We observe that $\theta_S$ starts at $16^{\circ}$ and increases up to the ideal mixing angle $35.264^{\circ}$. A different behavior is found for the angle $\theta_P$ that changes sign at $\rho_B\approx4\rho_0$. In fact, it starts at $-5.8^{\circ}$ and goes to the ideal mixing angle $35.264^{\circ}$, leading to a change of identity between $\eta$ and $\eta'$. We think this result might be a useful contribution for the understanding of the somewhat controversial question: under extreme conditions will the pion degenerate with $\eta$ or $\eta'$? 
The change of sign and the corresponding change of identity between $\eta$ and $\eta'$, effects that we do not observe in the finite temperature case, are due to differences in the behavior of the strange quark mass \cite{costa,costaB}.

\begin{figure}[t]
\hspace{0.3cm}\includegraphics[width=7cm,height=6.7cm]{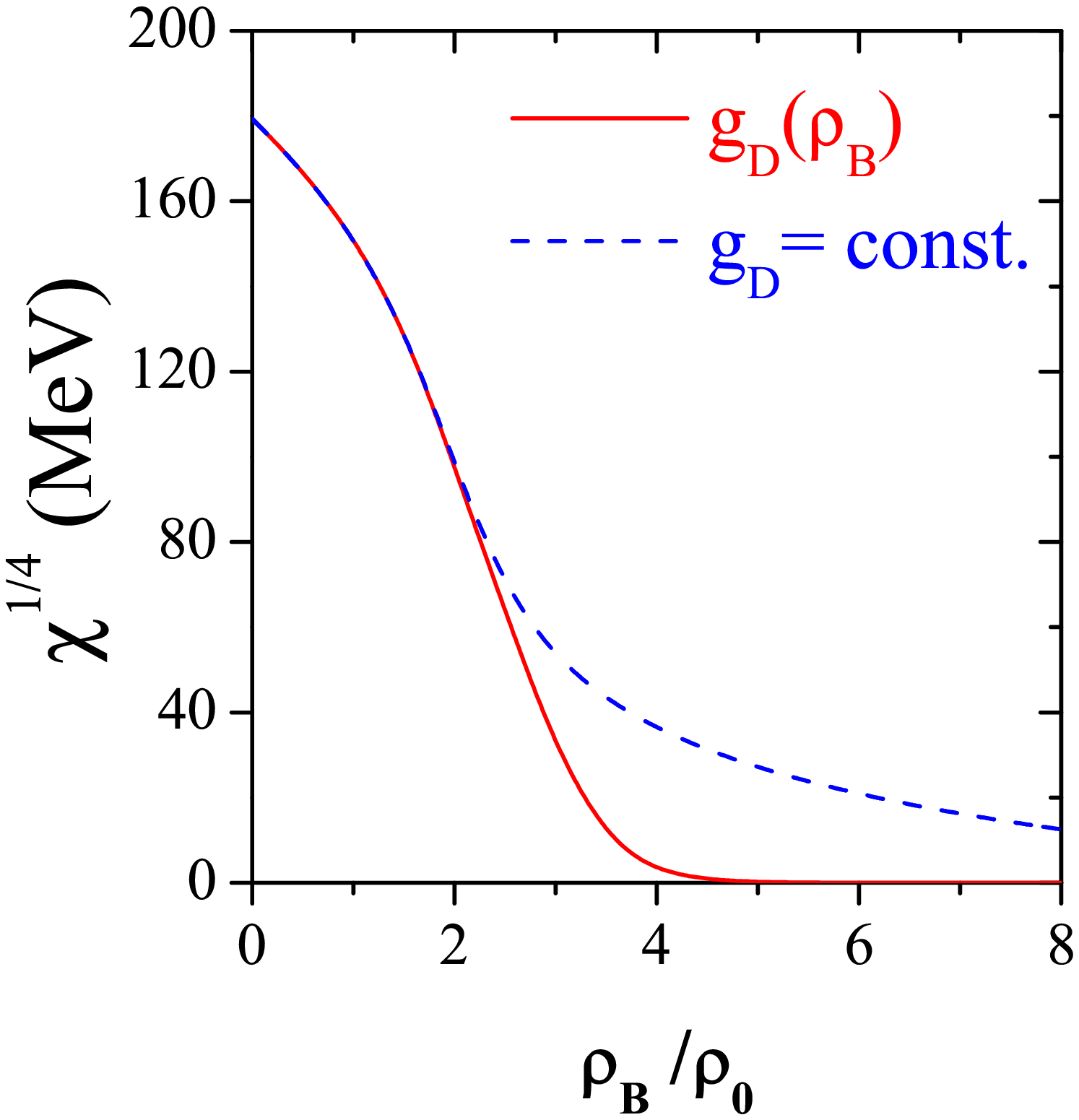}
\hspace{-1cm}\includegraphics[width=10cm,height=7cm]{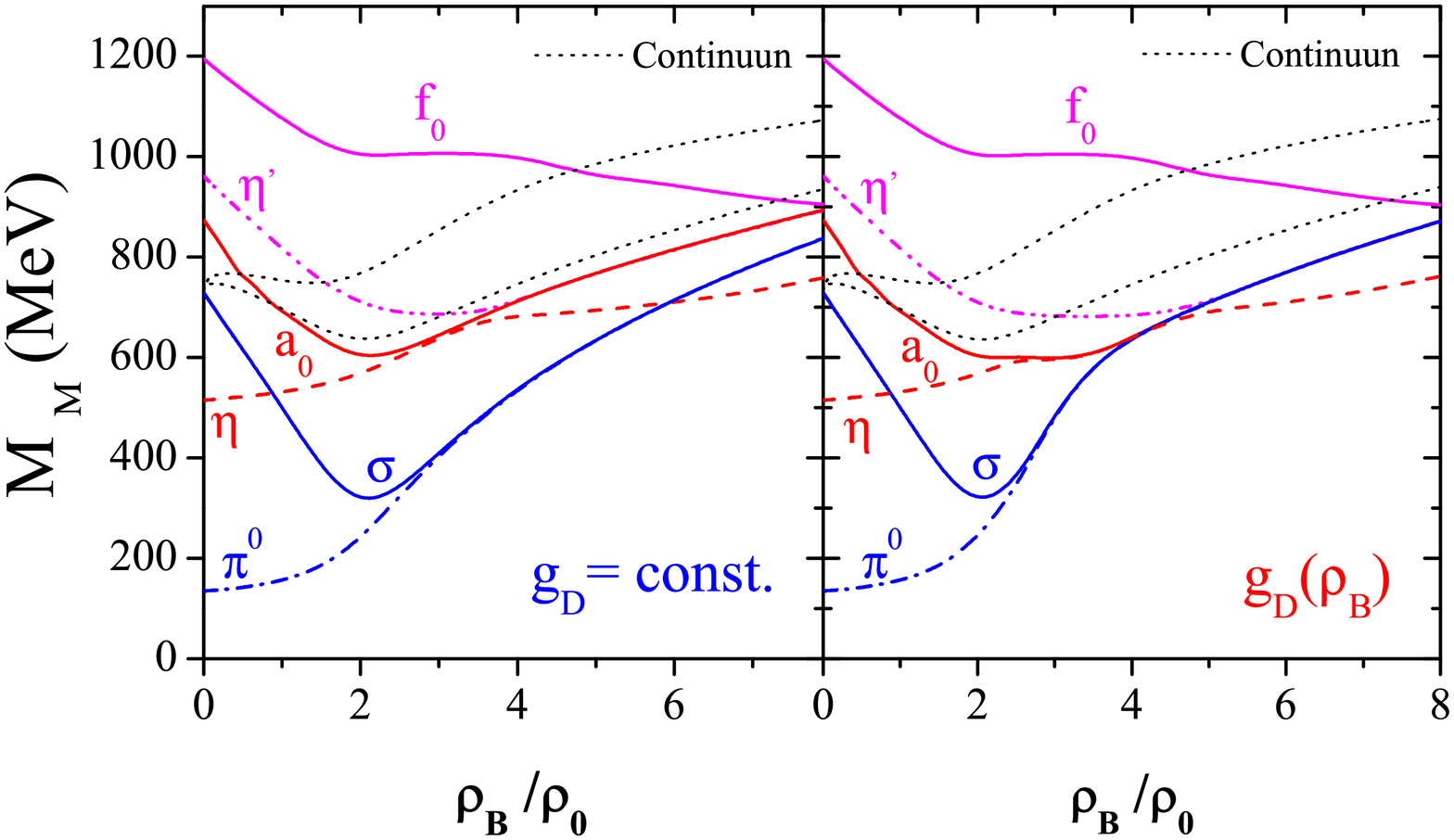}
\caption{Topological susceptibility (left panel): the solid (dashed) line represents our fitting with constant (density dependent) $g_D$. Meson masses, as functions of density, with $g_D$ constant (middle panel) and $g_D (\rho_B)$ (rigth panel). The dotted lines indicate the density dependence of the limits of the Dirac sea continua, defining $q\bar q$ thresholds for $a_0$ and $\eta^\prime$ mesons. }
\label{fig:dens}
\end{figure}

\begin{figure}[t]
\includegraphics[width=0.6\textwidth]{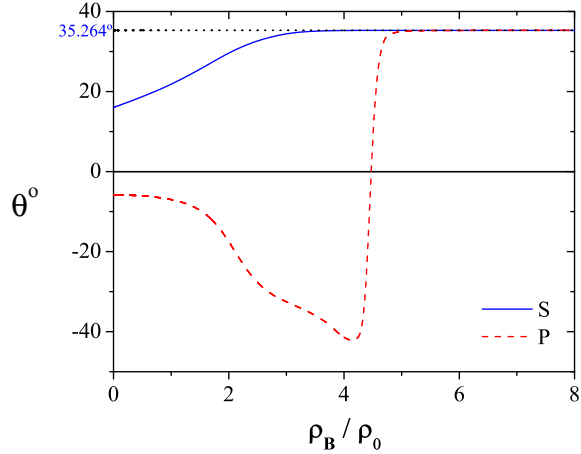}
\caption{The pseudoscalar and scalar mixing angles as functions of the baryonic density.}
\label{fig:ang}
\end{figure}

The meson masses, as function of the density, are plotted in Fig. 3, right panel. The results for constant $g_D$ are also presented (middle panel) for 
comparison purposes.
The SU(2) chiral partners ($\pi^0,\sigma$) are  bound states and become degenerate at $\rho_B=3\rho_0$.
With respect to the SU(2) chiral partners ($\eta,a_0$), the $a_0$ meson is 
always a purely non-strange quark system. For $\rho_B<0.8\rho_0$ $a_0$ is 
above the continuum and, when $\rho_B\geq0.8\rho_0$, $a_0$ becomes a bound 
state. At $\rho_B = 0$, the $\eta$ has a strange component and, as the density increases, $\eta$ becomes degenerate with $a_0$ at $4.0\rho_0\leq\rho_B\leq4.8\rho_0$ as expected. In this range of densities 
($\eta,a_0$) and ($\pi^0,\sigma$) are all degenerate. Suddenly, the $\eta$ 
mass separates from the others becoming a purely strange state. This is due 
to the behavior of  $\theta_P$ that changes  sign and goes to $35.264^{\circ}$ at $\rho_B\approx4.8\rho_0$. On the other hand, the $\eta'$, that starts as an unbound state and becomes bounded at $\rho_B>3.0\rho_0$, turns into a purely light quark system and degenerates with $\pi^0$, $\sigma$ and $a_0$ mesons \cite{costaUA1}.

Finally  we analyze  the behavior of charged mesons with density, plotted in Fig. 5. A glance at   the figure immediately reveals that:  the  chiral partners 
$(\pi^+\,, a_0^{\,+})$ and  $(\pi^-\,, a_0^{\,-})$, upper panel, become degenerate for  $\rho_B\approx 4 \rho_0$; the chiral partners $(K^+\,, \kappa^+), (K^-\,, \kappa^-)$, middle panel, and  $(K^0\,, \kappa^0), (\bar K^0\,, \bar\kappa^0)$, lower panel, do not degenerate in the region of densities considered. 
We notice that,  while the results for  $(\pi^\pm\,, a_0^{\,\pm})$  are affected by the dependence of $g_D$ on density, we find no substantial differences for the kaonic mesons, whether  $g_D$ is constant or not. In order to understand this,  let us remember that the effects of U$_A$(1) symmetry breaking or restoration appear explicitly, in the gap equations and in the meson propagators,  through the anomaly coupling, $g_D$, times a quark condensate.  For the pion and the $a_0$ propagators, the dependence on   the anomaly  enters through $g_D\, <\bar q_s q_s>$ so, with  $g_D$ a decreasing function of the density, this term will affect less and less the mesons masses as the density increases. Then, the convergence of the mesons reflects the restoration of  the U$_A$(1) symmetry. Since for kaonic mesons the  propagators depend on the anomaly through $g_D\, <\bar q_u q_u>  (<\bar q_d q_d> )$, the anomaly has little effect on the kaonic masses, as the density increases, whether $g_D$ is constant or not,  due to the strong decrease of the mass of the non-strange quarks. The dominant factor for the calculation of the masses of those mesons is the mass of the strange quark, which, although decreasing, remains  always very high.
We can   say that the restoration of the axial anomaly does not influence the behavior of kaons and of its chiral partners. 
In addition, we remark that $M_u$ and $M_d$ have different behaviors in neutron matter \cite{costa}, and the chiral asymmetry $\chi_A=|M_u-M_d|/(M_u+M_d)$ is always different from zero, even for high densities.    

\begin{figure}[t]
\includegraphics[height=.6\textheight]{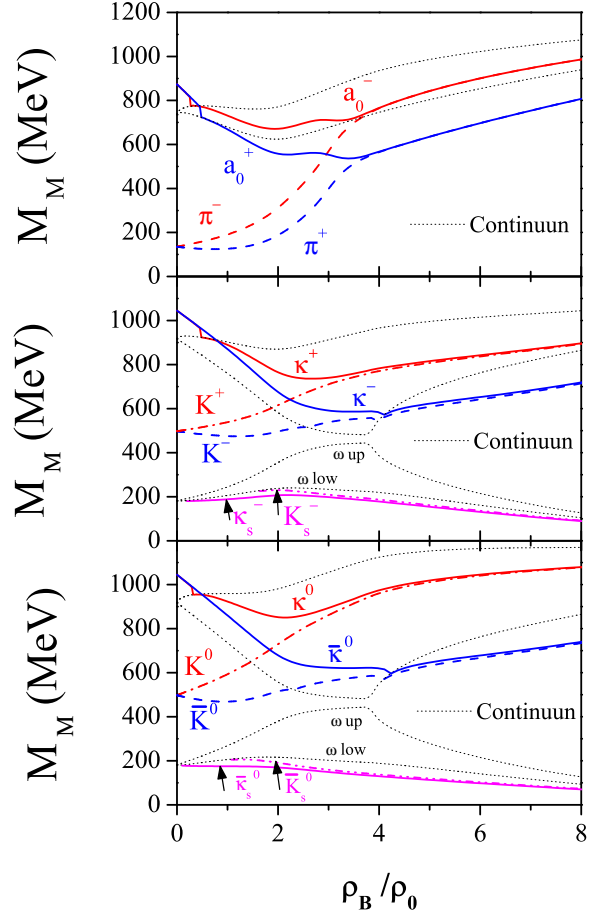}
\caption{ Meson masses, as functions of density, with  $g_D (\rho_B)$. The dotted lines indicate the density dependence of the limits of the Dirac   and the Fermi  ($\omega_{\rm low}$ and $\omega_{\rm up}$) sea continua, defining $q\bar q$ thresholds for the mesons. The low-lying mesonic solutions ( denoted by the subscript $S$) are also included.}
\label{fig:charged}
\end{figure}
 
We also  analyze  the low-lying  solutions, which we denoted by the subscript $S$. Below the lower limit ($\omega_{\rm low}$, in Fig. 5) of the Fermi sea continuum of particle-hole excitations  there are low bound states with quantum numbers of $K^{-}\,,\bar{K}^{0}\,$ and $\pi^+$. These  collective particle-hole excitations of the Fermi sea, associated to a first-order phase transition,  are manifestations of the violation of the isospin symmetry by the environment conditions. It is  interesting to look for the respective chiral partners and observe the behavior of the mass  splitting between scalar and pseudoscalar mesons as the   densities increases. The decreasing of this splitting for $K^{-}$ and $\bar{K}^{0}$ mesons can be seen in Fig. 5.

We notice that the convergence between  the different chiral partners always  occurs at densities where the mesons are bound states (see Fig. 5), i.e., they are collective excitations defined below the respective $q\bar q$ threshold.   


\section{Conclusions}
\label{Conc}

In this work we considered, via the U$_A$(1) anomaly, an explicit breaking of the axial symmetry in the vacuum state. The subsequent restoration of axial symmetry at non-zero temperature (density) has been then discussed using a temperature (density) dependent  anomaly coefficient. The case with $g_D={\rm cte}$ for all temperatures (densities) is also discussed.  We verified that in this case there is always an amount of U$_A$(1) symmetry breaking in the particle spectrum even when chiral symmetry restoration in the non-strange sector occurs at high temperature (density).  

Since  in all cases the chiral symmetry is explicitly broken by the presence of non-zero current quark mass terms, the chiral symmetry has been  realized through parity doubling rather than by massless quarks.
So, the identification of chiral partners and the study of its convergence is the criteria to study the effective restoration of chiral and axial symmetries. An important information is also provided by the mixing angles and we verify that, in the scenario of effective restoration of   axial  symmetry, the mixing angles  converge to  situations of ideal flavor mixing: (i) the $\sigma$ and $\eta$  mesons are pure non-strange $q\bar q$ states, while $f_0$ and $\eta^\prime$ are pure strange $s\bar s$ excitations for non-zero temperature case; (ii) the $\eta$ and $\eta^\prime$ change identities for neutron matter case.

In the conditions of explicit breaking of chiral symmetry  we worked, SU(3) symmetry is not exact and the strange sector does contribute with significant effects, even at high temperature (density) as it is visible in the behavior of $f_0$ and $\eta$ ($\eta^\prime$) mesons.

We can conclude that  the U$_A$(1) symmetry is, at least partially, restored above  the critical transition temperature of the SU(2) chiral phase transition. But, in the region of temperatures (densities) studied we do not observe signs indicating a full restoration of U(3)$\otimes$U(3) symmetry.
In fact,  as we work in a real world scenario ($m_u=m_d<<m_s$),  we only observe the return to symmetries of the classical QCD Lagrangian in the non--strange sector.
  
\begin{theacknowledgments}
Work supported by grant SFRH/BD/3296/2000 (P. Costa), CFT and by FEDER/FCT under project POCTI/FIS/451/94. 
\end{theacknowledgments}


\end{document}